\documentclass[12pt,a4]{article} 
\usepackage[koi8-r]{inputenc}
\font\mnf = msbm10 scaled \magstep1 %at 14pt}

\topmargin - 24pt
\begin{document}

%\newpage
\pagenumbering{arabic}
\thispagestyle{empty}

\begin{center}
M.V. LOMONOSOV MOSCOW STATE UNIVERSITY\\[2mm]
D.V. SKOBELTSYN INSTITUTE OF NUCLEAR PHYSICS\\[2mm]
119899 Moscow, Russia
\end{center}

\vspace{4cm}

%\begin{flushright}
%                 {\bf Preprint NPI MSU 00-$\phantom{------}$}
%\end{flushright}
%
%\vspace{1cm}

\begin{center}
{\Large\bf ONE-DIMENSIONAL SCATTERING PROBLEM,  \\[1mm]
            SELF-ADJOINT EXTENSIONS, \\[1mm]
          RENORMALIZATIONS AND POINT    \\[4mm]
            $\delta$-INTERACTIONS FOR COULOMB POTENTIAL} \\    %\\[2mm]

\vspace{2cm} 
{\large\bf V.S. Mineev} %\footnote{E-mail:
%                                         mineev@theory.npi.msu.su}} 

\vspace{3cm}
 {\bf Preprint NPI MSU 2000-24/628 \\
                         (revised version) }
\vfill

Moscow 2002

\end{center}
%%%%%%%%%%%%%%%%%%%%%%%%%%%%%

\newpage
\thispagestyle{empty}

$\left. \right.$

\vfill

{\noindent
{\bf V.S. Mineev} \\
E-mail address: \\
mineev@theory.sinp.msu.ru}\\

\vspace{1cm}

\begin{center}
              PREPRINT NPI MSU 2000-24/628 \\
                         (revised version) \\

\vspace{5mm}

%\noindent
{\bf One-dimensional scattering problem, self-adjoint extensions, 
renormalizations  and
point $\delta$-interactions for Coulomb potential}\\

\end{center}

\vspace{1cm}

{\bf ABSTRACT.}

In the paper the one-dimensional one-center scattering problem with the
initial potential $\alpha |x|^{-1} $ on the whole axis 
is treated and reduced to the search for allowable self-adjoint extensions.
 Using the laws of conservation as necessary conditions
 in the singular point alongside
with account of the analytical structure of fundamental solutions, it 
allows us to receive {\it exact} expressions for the 
wave functions (i.e. for the boundary conditions),
 scattering coefficients and the singular corrections to the potential,
as well as the corresponding bound state spectrum. It turns out
 that the point $\delta$-shaped correction
to the potential should be present {\it without fail} 
at any choice of the allowable
self-adjoint extension, moreover a form of these corrections
corresponds to the form of renormalization terms
obtained in quantum electrodynamics.

Thus, the proposed method shows the unequivocal connection among the boundary
conditions, scattering coefficients and $\delta$-shaped additions to the
potential. Taken as a whole, the method demonstrates the opportunities which
arise at the analysis of the self-adjoint extensions of the appropriate
Hamilton operator. And as it concerns the renormalization
 theory, the method can be treated
as a generalization of the Bogoliubov, Parasiuk and Hepp method of
renormalizations.

\vspace{1cm}

{\noindent
{\it AMS classification scheme numbers}: 34L40, 47A75, 46N50, 46F10, 46F20, 81T16 \\

\vspace{0,5cm}


\vfill

\begin{flushright}
                    \copyright V.S. Mineev, 2002
\end{flushright}

%%%%%%%%%%%%%%%%%%%%%%%%%%%%%%%%%%%%%%%%%%%%%%%%%%%%%%%%%%%%%%%%%%%%%%%%%%

\newpage
$\left. \right.$

\vskip 2cm

\section{Introduction}

Study of the singular interactions in the quantum theory has got already
rather long history. One of the first papers devoted to singular
 potentials is Ref. \cite {case-50}, and the study of the
 Schr{\"o}dinger equation with
$\delta$-shaped singularities ascends to Ref. \cite {kron-31}. In subsequent
years this direction of research caused extensive literature
 (see \cite{frank-71}, \cite {dem-75}),
 moreover a serious mathematical analysis of
arising in this connection problems was carried out rather recently (see \cite
{aghh-91}). In particular, a study of the three-dimensional
Schr{\"o}dinger equation was comparatively recently carried out,
 in which there are presented simultaneously the Coulomb and
point interactions, and moreover the Coulomb interaction being treated 
as perturbation
of the one-center three-dimensional problem \cite {zorb-80}, \cite {albev-83}
(the basis of approach, developed in Refs. \cite {zorb-80}, \cite
{albev-83}, was put down already in Ref. \cite {rel-43}). 

In the absolutely overwhelming majority of papers (as well as in Refs. \cite
{zorb-80}, \cite {albev-83}) precisely the three-dimensional case 
was examined, and  the partial expansion of wave functions 
(or the appropriate space $L^2
(\mbox {\mnf R}^3)$) in respect to an angular momentum was
carried out. In turn, the reduced
radial wave function was unequivocally determined by the
 boundary condition $\psi(0) = 0$, that correspond to the existence 
 of one-parameter family of the self-adjoint
operators. In addition, the angular component of
wave function was studied in detail.

It should be noted that for the first time the definition of the 
Hamiltonian  as a
self-adjoint operator in space $L^2 ({\mbox{\mnf R}}^3)$ was given
 in the paper \cite{bf-61}.

The one-dimensional case is investigated much less then the 
three-dimensional one. In the one-dimensional case, in contrast
with the two and three dimensional ones, it is
necessary to examine both linearly independent solutions
 that at the presence of
singularity result in the existence of four-parameter family of self-adjoint
extensions in space $L^2 (\mbox {\mnf R}) $. And it, in turn, can lead to an
opportunity of existence of additional point interactions.

In the present paper the one-dimensional one-center scattering problem on the
whole axis with the initial potential $\alpha |x|^{-1}$ is considered. The
generic solution of the stationary Schr\"odinger equation with the Coulomb 
potential is known, and the problem is reduced to a search for
 self-adjoint extensions, or, in the
narrower interpretation, to a search for boundary conditions 
at the singularity point,
satisfying the general principles of quantum theory. 

Incidentally, we shall note here that enormous number of the papers is 
devoted to various aspects of Coulomb interaction. However the number of the papers, 
devoted to the mathematical problems which are related to the presence of a 
non-itegrable singularity at a point of interaction, is few in number. In the 
present paper we shall restrict our consideration to the review of the papers 
of the general type relating to the problems under consideration only. The more 
comprehensive review of the latest papers, devoted to the given subject, will 
be cited in the following publication. We 
shall add here only, that in overwhelming majority of papers, dedicated to the 
one-dimensional Coulomb problem, the bound states and their eigenfunctions have 
been examined only. Very nearly unique exception (beeng devoted to somewhat 
different, Coulomb-type singular interaction) is the paper \cite{mosh93}, 
provoked objections given in the paper \cite{newt94} (the answer to these of 
objections see in the paper \cite{mosh94}).

For solving the specified problem of a choice 
of boundary conditions (i.e.,
for the concrete definition of parameter values defining, in turn,
 the appropriate
self-adjoint extension) the conditions of probability and flow
 conservation at the singularity point $x=0$ are
used in the paper as necessary conditions. Such approach 
in combination with the
method of variation of constants, the use of 
analytical structure of wave functions
and the theory of distributions allows  to solve completely the one-dimensional
scattering problem for the Coulomb potential. In this connection it appears
that the specified conditions {\it without fail} lead 
to the necessity of inclusion
in the potential the singular $\delta$-shaped additions.

It is necessary to note that the method used
 in the paper admits the opportunity
of existence of a lot of sets of the specified parameters, 
i.e. there is
ambiguity in a choice of allowable solutions and Hamiltonians. Such ambiguity
in the language of operator theory corresponds to  that the considered
unlimited symmetric (i.e. Hermitian) operator is not self-adjoint one
on the whole axis and permits
various self-adjoint extensions (see \cite {rs-1}, \cite {rs-2}).
Probably, for the first time the attention to the
fact of admissibility of several types of solutions was paid
in Refs. \cite {ll-63}, \cite {lieb-63}, which, in
turn, have resulted in the analysis of connection between 
the topology and quantum
statistics, and, accordingly, to the study of dependence  of the description of
bosons, fermions and paraparticles on the choice of boundary conditions in the
singularity point (see, e.c., \cite {abs-90}, \cite {bala-90}).

\section{Basic equations}

We shall consider the stationary one-dimensional Schr{\"o}dinger equation
$$
\psi''(x) + \left [{k^2} - u (x) \right] \psi (x) = 0 \eqno {(1)}
$$
with the potential
$$
u(x) =\varepsilon (x) \, {\alpha \over x} \,,
        \qquad x, \ \alpha \in {\mbox {\mnf R}} \,, \eqno {(2)}
$$
where $\varepsilon (x) $ is the sign function
$$
\varepsilon (x) =\theta (x) - \theta (-x) \eqno {(3)}
$$
($\theta (x) $ is the characteristic Heaviside stepwise function). We shall
construct a generic solution of the equation (1) using the Whittaker functions
appropriate to the scattering problem, i.e. we shall choose the solutions as the
fundamental pair (see Appendix)
$$
\psi_ + ^ {+, W} (k, x) = e^ {{\pi \alpha} \over {4k}} \,
 W_ {- {i \alpha \over 2k}, {1 \over 2}} (-2ikx)\, ,           \eqno {(4a)}
$$
$$
\psi_ + ^ {-, V} (k, x) = - e^ {{\pi \alpha} \over {4k}} \,
V_ {- {i \alpha \over 2k}, {1 \over 2}} (-2ikx)                \eqno {(4b)}
$$
for the positive half-axis and
$$
\psi_-^ {+, W} (k, x) = e^ {- {\pi \alpha \over 4k}} \,
W_ {{i \alpha \over 2k}, {1 \over 2}} (-2ikx)\, ,              \eqno {(5a)}
$$
$$
\psi_-^ {-, V} (k, x) =- e^ {- {\pi \alpha \over 4k}} \,
V_ {{i \alpha \over 2k}, {1 \over 2}} (-2ikx)                   \eqno {(5b)}
$$
for the negative half-axis. Functions (4), (5), as it is remarked in Appendix, 
are characterized by asymptotic behavior
$$
\lim_ {x \to + \infty} \psi_ + ^ {+, W} (k, x) =
e^ {i \left (kx - {\alpha \over 2k} \ln (2kx) \right)} 
\left [1 + {\rm O} \left (1 \over -2ikx \right) \right] \,,     \eqno {(6a)}
$$
$$
\lim_ {x \to + \infty} \psi_ + ^ {-, V} (k, x) =
 e^ {-i \left (kx - {\alpha \over 2k} \ln (2kx) \right)} 
\left [1 + {\rm O} \left (1 \over 2ikx \right) \right] \,,      \eqno {(6b)}
$$
$$
\lim_ {x \to -\infty} \psi_-^ {+, W} (k, x) =
 e^ {i \left (kx + {\alpha \over 2k} \ln (2kx) \right)} 
\left [1 + {\rm O} \left (1 \over -2ikx \right) \right] \,,     \eqno {(6c)}
$$
$$
\lim_ {x \to -\infty} \psi_-^ {-, V} (k, x) =
 e^ {-i \left (kx + {\alpha \over 2k} \ln (2kx) \right)} 
\left [1 + {\rm O} \left (1 \over 2ikx \right) \right] \,,      \eqno {(6d)}
$$
and their Wronskians (see Appendix) look like
$$
{\cal W} \left\{\psi_+^{+,W}(k,x), \ \psi_+^ {-,V}(k, x)\right\}=
             -2ik\,,                                             \eqno{(7)}
$$
$$
{ \cal W} \left\{\psi_-^ {+, W} (k, x), \ \psi_-^ {-, V} (k, x) \right\} =
             -2ik\,.                                            \eqno {(8)}
$$
The choice of solutions (4), (5) with the asymptotic behavior (6) 
fixes the behavior of solutions of the equation (1) on infinity.
 However, as it was mentioned above, the
Schr{\"o}dinger operator (1) with the singular potential (2)
 (as well as the momentum operator) is an example of the closed symmetric
but not self-adjoint operator
(see \cite {rs-1}, \cite {rs-2}), and it can have in the examined case even not
four but eight-parametrical family of self-adjoint extensions (about the
four-parametrical families of self-adjoint extensions see \cite {carfa-90},
\cite {car-91}, and also \cite {aghh-91}). 
The particular choice of the
self-adjoint extension, as it is known \cite {rs-2}, defines appropriate
physics. In turn, from the point of view of physics, for definition of a
complete set of states it is necessary to fix 
a behavior of wave functions (and operators) near the singular points
 that define the extension of the operator up to the self-adjoint one.

Before fixing the behavior of solutions in zero we remind the standard
definition of the reflection and transmission coefficients. Usually in the
elementary cases, corresponding to ``good'' potentials
(see., e.g., \cite{ber-83})
 (i.e. in the cases when the operator is self-adjoint or, 
in other words, its
deficiency indices are equal to $(0, 0)$ (see \cite {bf-61},
and also \cite {rs-2}, \cite
{aghh-91})), one takes as a fundamental pair the pair
 of fundamental solutions (or the
Iost solutions) $f_1 (k, x) $ and $f_1 (-k, x) $ described by the asymptotic
behavior:
$$
\lim_ {x \to + \infty} {\left [e^ {-ikx} \, f_1 (k, x) \right]} =1 \,,
                                                                \eqno{(9a)}
$$
or $f_2 (k, x) $ and $f_2 (-k, x) $ with the behavior like
$$
\lim_ {x \to -\infty} {\left [e^ {ikx} \, f_2 (k, x) \right]} =1 \,.
                                                                \eqno{(9b)}
$$
As any third solution is a combination of two linearly independent solutions,
it is possible to write that, for example,
$$
f_2 (k, x) =c_ {11} (k) \, f_1 (k, x) + c_ {12} (k) \, f_1 (-k, x) \,,
                                                               \eqno{(10a)}
$$
$$
f_1 (k, x) =c_ {21} (k) \, f_2 (-k, x) + c_ {22} (k) \, f_2 (k, x) \,.
                                                              \eqno{(10b)}
$$
Considering now the limiting values of the relations (10) it is
easy to see that Eq.
(10a) represents a solution of the equation (1), which at $x \to - \infty$
approaches $e^ {-ikx} $, and at $x \to + \infty$ is the
linear combination $c_{11} e^ {ikx} + c_ {12} e^ {-ikx} $.
 Thus, the solution (10a) corresponds to the
scattering problem, in which the wave with the amplitude 
$c_ {12}$  falls on the scattering
potential from $x = + \infty$ (the R-case), and it is partially reflected
with the amplitude $c_ {11}$, and partially passes to $- \infty$ 
with the amplitude equal to $1$. It means that the reflection and 
transmission coefficients are of the
form:
$$
R_R (k) = {c_ {11} (k) \over c_ {12} (k)} \,,                \eqno {(11a)}
$$
$$
T_R (k) = {1 \over c_ {12} (k)} \,. \eqno {(11b)}
$$
Similarly, the solution (10b) describes the scattering problem with the wave
falling from the left (L-case) and with the coefficients
$$
R_L (k) = {c_ {22} (k) \over c_ {21} (k)} \,,                \eqno {(12a)}
$$
$$
T_L (k) = {1 \over c_ {21} (k)} \,.                          \eqno {(12b)}
$$

In the case examined in the present paper the realization
 of such the program has that
difficulty that the Whittaker functions (a6) and (a7) have the regular
singular point at $z=0$ , being the logarithmic branch point,
 i.e. the connection of
functions $f_1 (k, x) $ and $f_1 (-k, x) $ (and, accordingly,
 $f_2 (k, x) $ and
$f_2 (-k, x) $) is ambiguous, and we do not know {\it a priori},
 at which sheet
the reflected and transmitted waves will appear. However, the fact that the
analytical continuation of any solution of the equation (1) exists
(see Appendix) and both functions $W_ {p, m} (z\, e^ {2 i \pi s})$ and
 $V_ {p, m} (z\, e^ {2i \pi s})$ are solutions
 of the equation (1), allows to present any solution of the
Schr{\"o}dinger equation (1)  as their linear combination.
 In our case it means that at $x > 0$ the expression
$$
\psi_ + (k, x) =\alpha_2^ + (k) \, \psi_ + ^ {-, V} (k, x) +
 \beta_2^ + (k) \, \psi_ + ^ {-, V} (e^ {2i \pi s} \, k, x)    \eqno {(13)}
$$
is valid. Accordingly, on the negative half-axis
$$
\psi_- (k, x) =\alpha_2^- (k) \, \psi_-^ {-, V} (k, e^ {2i \pi r} \, x) +
\beta_2^- (k) \, \psi_-^ {-, V} (e^ {2i \pi s} \, k, e^ {2i \pi r} \, x) \,.
                                                                \eqno {(14)}
$$
In addition, it is necessary to take into account
that in the expressions (13), (14) the passage 
of the variable $x$ through the singular point $x=0$ can,
generally speaking, change factors at the appropriate wave functions (this
phenomenon is connected with the indices of the regular singular point, see
Appendix, and also \cite {olv-90}). Introducing the designations
$$
\alpha_2^- (k) =\alpha_2^ + (k) \, Q_1=\alpha_2 (k) \, Q_1 \,,   \eqno {(15)}
$$
$$
\beta_2^- (k) =\beta_2^ + (k) \, Q_2 =\beta_2 (k) \, Q_2 \,,     \eqno {(16)}
$$
it is possible to write down the solution of the equation (1) on the 
entire line as
$$
  \psi (x) = \alpha_ {2} (k) \left[\psi_ + ^ {-, V} (k, x) \, \theta (x) +
     Q_1\, \psi_-^ {-, V} (k, e^ {2i \pi r} \, x) \, \theta (-x) \right] + $$
   $$ \beta_2 (k) \, \left[\psi_+^{-,V} (e^{2i \pi s} \, k, x) \,
 \theta(x) +
     Q_2\, \psi_-^ {-, V} (e^ {2i \pi s} \, k, e^ {2i \pi r} \, x) \,
           \theta (-x) \right] \,.                              \eqno {(17)}
$$

The formulas of analytical continuation of the Whittaker functions (a30) can be
submitted as
$$
V_ {p, {1 \over 2}} \left(e^ {2i \pi s} \, z \right) =
    b_s^V (p) \, V_ {p, {1 \over 2}} (z) +
    b_s^W (p) \, W_ {p, {1 \over 2}} (z)\,,
                                                                \eqno {(18)}
$$
where
$$
b_s^V (p) = -s\, e^ {2i \pi p} + (s + 1) \,,                    \eqno {(19)}
$$
$$
b_s^W (p) =-s\, {2i \pi\, e^ {i \pi p} \over \Gamma (p) \,\Gamma (1 + p)} \,.
                                                                \eqno {(20)}
$$
It allows to rewrite the expression (17) in the form
$$
  \psi (x) = \left[\alpha_ + ^- (k) \, \psi_ + ^ {-, V} (k, x) +
     \alpha_ + ^ + (k) \, \psi_ + ^ {+, W} (k, x) \right] \theta (x) + $$
   $$ \left[\alpha_-^- (k) \, \psi_-^ {-, V} (k, x) +
      \alpha_-^ + (k) \, \psi_-^ {+, W} (k, x) \right] \theta (-x) \,,
                                                                \eqno{(21)}
$$
where
$$
\alpha_ + ^- (k) = \alpha_2 (k) + \beta_2 (k) \, 
       b_s^V \left(- {i\alpha \over 2k} \right) \,,            \eqno {(22)}
$$
$$
\alpha_ + ^ + (k) = -\beta_2 (k) \, 
       b_s^W \left(- {i\alpha \over 2k} \right) \,,           \eqno {(23)}
$$
$$
\alpha_-^-(k) = Q_1\,\alpha_2(k) \, b_r^V \left({i\alpha \over 2k} \right)+
 Q_2\,\beta_2 (k)\,b_{r + s} ^V \left({i\alpha \over 2k} \right) \,, 
                                                               \eqno{(24)}
$$
$$
\alpha_-^ + (k) = -\left[Q_1\, \alpha_2 (k) \,
 b_r^W \left({i\alpha \over 2k} \right) 
+ Q_2\, \beta_2 (k) \, b_ {r + s} ^W \left({i\alpha \over 2k} \right) \right]
\,. 
                                                               \eqno{(25)}
$$
We should note that although the expression (17) corresponds to the formula
(10b) it has completely general character, the wave function $\psi (x) $ (21)
represents an arbitrary solution of the equation (1) with the 
potential (2). The appropriate representation can be obtained at the 
construction of analogue of the formula (10a).

We return now to the scattering problem 
briefly described by the relations (9)--(12).
 Let, for example, $\psi (x) =f_2 (k, x) $ be a fundamental solution with the
asymptotic behavior similar to the behavior (9b).
Comparison of the expressions (21) and (10a) shows that the solution
 describing the  scattering of particles
incident from the right (R-case) can be presented as
$$
\psi (x) =f_2 (k, x) = $$
       $$ {\cal A} _R (k) \, \psi_ + ^ {-, V} (k, x) \, \theta (x)+
     {\cal B} _R (k) \, \psi_ + ^ {+, W} (k, x) \, \theta (x) +
      \psi_-^ {-, V} (k, x) \, \theta (-x) \,,               \eqno {(26)}
$$
and the expressions for the scattering coefficients (11) in this case get the
form
$$
R_R (k) = {{\cal B} _R (k) \over {\cal A} _R (k)} \,,        \eqno {(27)}
$$
$$
T_R (k) = {1 \over {\cal A} _R (k)} \,.                      \eqno {(28)}
$$

We shall use now a rather modified method of variation of constants, i.e. we
shall treat the factors $\alpha_2 (k) $ and $\beta_2 (k) $
 in the expression (17) (and,
accordingly, $\alpha_ {+} ^ {-} (k) $, $\alpha_ {+} ^ {+} (k) $,
 $\alpha_ {-} ^{-} (k) $,
 $\alpha_ {-} ^ {+} (k) $ in the formulas (22)--(25)) as values
depending on $x$. If one presents the expressions for
 $\alpha_2 (k, x) $ and $\beta_2(k, x) $ as
$$
\alpha_ {2} (k, x) = A_1 (k) \, \theta (x) + A_2 (k) \, \theta (-x) \,,
                                                              \eqno{(29)}
$$
$$
\beta_ {2} (k, x) = B_1 (k) \, \theta (x) + B_2 (k) \, \theta (-x) \,,
                                                              \eqno{(30)}
$$
the comparison between Eqs. (17)--(25) and Eq. (26) 
yields in the examined case of particle
incident from the right the following expression for
$A_2(k)$ and $B_2(k)$:
$$
A_2 (k) = {1 \over Q_1} \, {r + s \over s} \,,                \eqno {(31)}
$$
$$
B_2 (k) =- {1 \over Q_2} \, {r \over s} \,,                   \eqno {(32)}
$$
i.e.
$$
{ \cal A} _R (k) =A_1 (k) + B_1 (k) \, b_s^V \left(- {i\alpha \over 2k}
\right) \,,
                                                              \eqno{(33)}
$$
$$
{ \cal B} _R (k) =- B_1 (k) \, b_s^W \left(- {i\alpha \over 2k} \right) \,,
                                                             \eqno{(34)}
$$
and
$$
\alpha_ {2} (k, x) = \left[{\cal A} _R (k) +
 {\cal B} _R (k) \, {b_s^V \left(- {i\alpha \over 2k} \right) \over 
 b_s^W \left(- {i\alpha \over 2k} \right)} \right] \theta (x) + 
   {1 \over Q_1} \, {r + s \over s} \, \theta (-x) \,,       \eqno {(35)}
$$
$$
\beta_ {2} (k, x) =
- {\cal B} _R (k) \, {1 \over b_s^W \left(- {i\alpha \over 2k} \right)} \,
        \theta (x) -
{ 1 \over Q_2} \, {r \over s} \, \theta (-x) \,.              \eqno {(36)}
$$
The expressions (35), (36) obviously demonstrate the difference in
definition of the
scattering coefficients in the treated here singular and 
elementary standard cases
(compare the formulas (10)--(12) and (17), (35)--(36), (27)--(28)).

Usual for the method of constant variation condition of the absence of terms
related to the differentiation of wave function coefficients
 leads in our case to the requirement of absence of a term
$$
 \left[{\cal A} _R (k) \, \psi_ + ^ {-, V} (k, x) +
     {\cal B} _R (k) \, \psi_ + ^ {+, W} (k, x) -
      \psi_-^ {-, V} (k, x) \right] \delta (x) \,            \eqno {(37)}
$$
in the derivative ${f_2}'(k, x)$.
Expansion of the functions $\psi_ {+} ^ {-, V} (k, x)$,
 $\psi_ {+} ^ {+, W} (k, x)
$ (and $\psi_ {-} ^ {-, V} (k, x) $) (a22) and (a24) with the account of the
operator relation $x\,\delta(x) =0$ and the expressions (33), (34) results in a
condition
$$
e^ {\pi \alpha \over 4k} \left[\left(A_1 (k)
 + B_1 (k) \, b_s^V \left(- {i\alpha \over 2k} \right) \right) 
 V_ {- {i \alpha \over 2k}, {1 \over 2}} (0) +
 B_1 (k) \, b_s^W \left(- {i\alpha \over 2k} \right) \,
 W_ {- {i \alpha \over 2k}, {1 \over 2}} (0) \right] - $$
     $$ e^ {- {\pi \alpha \over 4k}} \, 
 V_ {{i \alpha \over 2k}, {1 \over 2}} (0) = 0 \,.            \eqno {(38)}
$$
In this case
$$
 f'_2 (k, x) = \alpha_ {2} (k, x) \left[{\psi_+^{-,V}}'(k, x) \, \theta(x) +
     Q_1\, {\psi_-^ {-, V}}'(k, e^ {2i \pi r} \, x) \, \theta (-x) \right] +
$$
 $$ \beta_2(k,x) \left[{\psi_ + ^ {-, V}}'(e^{2i \pi s}\,k,x) \,\theta (x) +
     Q_2\, {\psi_-^ {-, V}}'(e^ {2i \pi s} \, k, e^ {2i \pi r} \, x) \,
          \theta (-x) \right] \,
                                                              . \eqno {(39)}
$$
Expression for the current density defined in the standard manner: 
$$
j(k, x) = {1 \over {2i}} \left[\psi^ * (k, x) \, \psi' (k, x) -
                    \psi (k, x) \, {\psi^*}'(k, x) \right] \,, \eqno {(40)}
$$
in view of the Wronskians (7), (8) and (a31)--(a33) takes the form
$$
j(k) =k \left(-{|{\cal A}_R(k)|}^2 + {|{\cal B}_R(k)|}^2 \right)\,,
                                                               \eqno {(41)}
$$
and the law of current conservation (i.e. the unitarity condition) is of the
form
$$
{ | {\cal A} _R (k) |} ^2 - {| {\cal B} _R (k) |} ^2 = 1 \,,    \eqno {(42)}
$$
or in the language of the scattering coefficients (27), (28):
$$
{ | R_R (k) |} ^2 + {| T_R (k) |} ^2 = 1 \,.                   \eqno {(43)}
$$
We note that the condition (38) coincides with the condition of even
continuation of the function $f_2 (k, x) $
 when passing through the point $x=0$, i.e.
the probability conservation in zero is provided.

Calculating the second derivative of the function $\psi (k, x) =f_2 (k, x) $ and
substituting the resultant expression into 
the equation (1) with the potential (2),
allowing for the expressions (17), (21), (26), we get
$$
 \left[{\cal A} _R (k) \, {\psi_ + ^ {-, V}}'(k, x) +
     {\cal B} _R (k) \, {\psi_ + ^ {+, W}}'(k, x) -
      {\psi_-^ {-, V}}'(k, x) \right] \delta (x) + $$
   $$ {\cal A} _R (k) \left\{{\psi_ + ^ {-, V}}''(k, x) +
             \left[k^2-\varepsilon (x) \, {\alpha \over x} \right] 
                           {\psi_ + ^ {-, V}} (k, x) \right\} \theta (x) + $$
   $$ {\cal B} _R (k) \left\{{\psi_ + ^ {+, W}}''(k, x) +
             \left[k^2-\varepsilon (x) \, {\alpha \over x} \right] 
                           {\psi_ + ^ {+, W}} (k, x) \right\} \theta (x) + $$
   $$ \left\{{\psi_-^ {-, V}}''(k, x) +
             \left[k^2-\varepsilon (x) \, {\alpha \over x} \right] 
                       {\psi_-^ {-, V}} (k, x) \right\} \theta (-x) =0 \, ,
                                                                \eqno{(44)}
$$
It is seen from this expression that the construction of a 
solution like Eq. (17), i.e. the
coordination of the solutions (13) on $\mbox {\mnf R} _ {+} $
 and (14) on $ {\mbox{\mnf R}}_{-} $ at zero,
 results in the occurrence of the additional first term
in Eq. (44). One should think that this term is possible to treat as a singular
addition to the potential, being a display of 
boundary conditions at zero. However
(and the subsequent calculation confirms this) in this case 
it is impossible to be sure
in the current conservation and the fulfillment of the
 condition (38). Therefore we
shall enter in Eq. (44) the mutually compensating each other terms of the form
$\alpha_2 (k, x) \, \psi_i (k, x) \,\delta V_i (k, x) $ and
 $\beta_2 (k, x) \, \psi_i (k, x) \,\delta V_i (k, x) $
 (where $\psi_i (k, x) =\psi_ {+} ^ {-, V} (k, x), \psi_ {+} ^ {+, W} (k, x),
\psi_ {-} ^ {-, V} (k, x) $,
 $\psi_ {-} ^ {+, W} (k, x) $). The occurrence of these terms means that the
functions 
$\psi_i (k, x) $ should satisfy the Schr{\"o}dinger equation in view of
boundary conditions, i.e. the equations for them take the form
$$
 {\psi_i}''(k, x) +
    \left[k^2-\varepsilon (x) \, {\alpha \over x} -\delta V_i (k, x) \right]
               \psi_i (k, x) =0 \,                            \eqno {(45)}
$$
(for some mathematical details of an opportunity of co-existence
 of the Coulomb  and
point potentials (specifically, as regards KLMN-theorem)
see \cite {aghh-91}, and also \cite {rs-2}).
We shall search for the singular additions $\delta V_i (k, x) $ as
$$
\delta V_i(k,x)=v_i\,{{\psi_i}'(k,x) \over \psi_i (k,x)} \, \delta (x)\, ,
                                                              \eqno {(46)}
$$
where the parameters $v_i=v_ {+} ^ {-, V} $, $v_ {-} ^ {-, V} $,
 $v_ {-} ^ {+, W}$, $v_ {+} ^ {+, W} $ for the appropriate $\psi_i (k, x) $.
On the one hand the form (46) for $\delta V_i (k, x) $ makes it 
to be  similar to
the first term of the formula (44), and on the other hand it
represents the singular
additions to the potential as the logarithmic derivative
 (see, e.g., \cite {bzp-71}).

Uniting the relations (45) and taking into account the expressions (33)--(36)
and (26) we obtain that in the case of particle incident from
the right the imposure of boundary conditions at $x=0$ leads to the appropriate
wave function $f_2 (k, x)$, which should satisfy the equation
$$
 {f_2}''(k, x) +
        \left[ k^2-\varepsilon (x) \, {\alpha \over x} -
 \delta V_ {f_2} (k,x) \right] f_2 (k, x) =0 \,,               \eqno {(47)}
$$
where
$$
\delta V_ {f_2} (k, x) = \left[
         {{{\cal A} _R (k) \, v_ + ^ {-, V} \, {\psi_ + ^ {-, V}}'(k, x) +
            {\cal B} _R (k) \, v_ + ^ {+, W} \, {\psi_ + ^ {+, W}}'(k, x)}
\over 
     {{\cal A} _R (k) \, {\psi_ + ^ {-, V}} (k, x) +
     {\cal B} _R (k) \, {\psi_ + ^ {+, W}} (k, x)}} \, \theta (x) + \right.
$$
        $$ \left. v_-^ {-, V} \, {{\psi_-^ {-, V}}'(k, x) \over
 \psi_-^ {-, V} (k, x)} \, \theta (-x) \right] \delta (x) \, , \eqno {(48)}
$$
under the condition of absence of the first term in the relation (44).
 In contrast with the condition (38) the structure of expansion
 of the functions $ {\psi_i}'(k, x)$ (a23)
 and (a25) with the account of the
operator relation $x\,\delta (x) =0$ leads to necessity of the
fulfillment of two conditions
$$
e^ {\pi \alpha \over 4k} \, \left[ \left( A_1 (k)
 + B_1 (k) \, b_s^V \left(- {i\alpha \over 2k} \right) \right)
        \left(1 + v_ + ^ {-, V} \right)
 V'_ {- {i \alpha \over 2k}, {1 \over 2}} (0) + \right. $$
    $$ \left. B_1 (k) \, b_s^W \left(- {i\alpha \over 2k} \right)
        \left( 1 + v_ + ^ {+, W} \right)
 W'_ {- {i \alpha \over 2k}, {1 \over 2}} (0) \right] - 
      e^ {- {\pi \alpha \over 4k}} \,
        \left( 1-v_-^ {-, V} \right) 
 V'_ {{i \alpha \over 2k}, {1 \over 2}} (0) = 0 \,,            \eqno {(49)}
$$
$$
e^ {\pi \alpha \over 4k} \left[ \left( A_1 (k)
 + B_1 (k) \, b_s^V \left(- {i\alpha \over 2k} \right) \right)
        \left(1 + v_ + ^ {-, V} \right)
 {1 \over {\Gamma \left(- {i\alpha \over 2k} \right)}} + \right. $$
    $$ \left. B_1 (k) \, b_s^W \left(- {i\alpha \over 2k} \right)
        \left(1 + v_ + ^ {+, W} \right)
   {1 \over {\Gamma \left ({i\alpha \over 2k} \right)}} \right] - 
      e^ {- {\pi \alpha \over 4k}} \,
        \left(1-v_-^ {-, V} \right)
   {1 \over {\Gamma \left( {i\alpha \over 2k} \right)}} = 0 \,. \eqno {(50)}
$$
The condition (50) provides the absence of terms of the form
 ${\ln x} \, \delta (x)$ in the expansion. 
 
 The fulfilled construction requires some comments.
 
 First of all it is necessary to note that 
 it is seen from the expression (44) and the
reasons bringing to the form (48) of the the singular addition 
to the potential, that apart from providing a
continuity of the function $f_2 (k, x)$
 when passing through zero (as it was specified above),
  the condition (38) is just 
the condition of the $\delta$-framed form of the singular addition
(48). Default of the condition (38), in particular,
 can lead to appearance  of terms proportional to $\delta'(x) $ in Eq. (1).

Taken as a whole, the fact of the presence of the singular
 addition (48) in Eq.
(47)  requires some explanation. The matter is that an appearance
of such the term is completely natural 
from the point of view of the theory of distributions.
And just the extension of the continuous linear functional (it is the
Hamiltonian in our case), defined at
$ {\cal J} ({\mbox {\mnf R}} \setminus 0)=\{f\in {\cal J} | {\rm supp} f
\subset 
{ \mbox {\mnf R}} \setminus 0 \}$, over the whole space 
${\cal J} ({\mbox {\mnf R}})$
 is referred to as the  ``renormalization'' for the considered functional
\cite {rs-1}. In addition, the renormalization of the functional has the
support $\{0\}$, i.e.,  generally speaking, represents an infinite series of
the form
$\sum_\alpha c_\alpha \, D^\alpha \, \delta (x)$, where $c_\alpha$ are some
free constants, and $D^\alpha$ is the derivative 
in the sense of distributions  (or the
weak derivative) (see \cite {rs-1}). Just so the renormalizations  were
interpreted in the theory of the $R$-operation of Bogoliubov, Parasiuk and
Hepp \cite {bp-57}, \cite {hepp-66} (see also \cite {bsh-73}). However in the
Bogoliubov, Parasiuk and Hepp theory the renormalizations
of terms of the Feynman expansion  were considered only,
 moreover the Pauli--Willars
regularization  was used for calculation of the constants $c_\alpha$ (which
played the role, for example,  of the mass and charge renormalizations) (see,
e.g., \cite {bsh-73}). In our case, as it was specified above, the whole
Hamiltonian is considered as the appropriate functional, and the
renormalization constants $c_\alpha$ are defined
 in a completely different way.
As one can see from the expression (48), the unique coefficient $c_0$ present
in the examined case is the function of parameters $A_i$, $B_i$ and $v_i$. In
turn, these parameters are defined by Eqs. (38), (49) and (50) received as a
result of the integration of expressions,
 which contain the products of distributions
$\theta (\pm x) \, \delta (x)$. Generally speaking, the product of
several distributions is not well defined integrable
function. Nevertheless, the product $\theta (\pm x) \, \delta (x)$ is defined
(for example, in sense of the improper limit transition, see \cite
{bsh-73}) for the basic functions continuous
 in the singular point (usually it
is natural to suppose for such basic functions $\theta (\pm x) \, \delta (x)
=1/2 \, \delta (x) $). In our case the basic function is not defined in the
singularity point  and the product of distributions $\theta (\pm x)$ and
$\delta (x)$ requires an extra-definition. In other words, the problem
of definition of the, generally speaking, arbitrary coefficients $c_ {\pm} $ in
the relations
 $\int f(x) \, \theta (\pm x) \, \delta (x) \, dx = c_ {\pm} \, f(0^
{\pm})$ arises. But Eqs. (38), (49) and (50) are just the equations determining the
the coefficients $c_\pm$, which are the combinations of parameters $A_i$,
$B_i$ and $v_i$. Thus, in solving the system of these equations we shall
define the products $\theta (\pm x) \, \delta (x)$, and this result,
 in turn, will
determine  the $\delta$-framed renormalization additions in the expression
(48).

\section{Solutions. Boundary conditions. Scattering coefficients. Bound states. 
Renormalizations}

Solving the system of the equations (38), (49), (50),
for example, in respect to $A_1(k) $, $B_1 (k) $ and $v_ {-} ^ {-, V} $
 and using the expressions (33), (34)
we get the generic solution:
$$
{ \cal A} _R (k) = {{4 \pi k \left(1 + v_ + ^ {+, W} \right)} \over 
       {\alpha \left(2 + v_ + ^ {-, V} + e^ {\pi \alpha
                          \over k} \, v_ + ^ {-, V} + 
       v_ + ^ {+, W} -
  e^ {\pi \alpha \over k} \, v_ + ^ {+, W} \right) \,
           \Gamma^2 \left( {i\alpha \over 2k} \right)}} \,,    \eqno {(51)}
$$
$$
{ \cal B} _R (k) = {{\left( 1 + e^ {\pi \alpha \over k}
              \right) \left(1 + v_ + ^ {-, V} \right)} \over 
       {e^ {\pi \alpha \over 2k} \, 
          \left(2 + v_ + ^ {-, V} +
        e^ {\pi \alpha \over k} \, v_ + ^ {-, V} + 
       v_ + ^ {+, W} - e^ {\pi \alpha \over k} \,
                            v_ + ^ {+, W} \right)}} \,,         \eqno {(52)}
$$
$$
v_-^ {-, V} = {{-4-3v_ + ^ {-, V} -
        e^ {\pi \alpha \over k} \, v_ + ^ {-, V} - 
       3v_ + ^ {+, W} + e^ {\pi \alpha \over k} \,
       v_ + ^ {+, W} -2 v_ + ^ {-, V} \, v_ + ^ {+, W}} \over  
       {-2-v_ + ^ {-, V} -
        e^ {\pi \alpha \over k} \, v_ + ^ {-, V} - 
       v_ + ^ {+, W} + e^ {\pi \alpha \over k} \,
                            v_ + ^ {+, W}}} \,.                 \eqno {(53)}
$$

At once the absence of the factors of the
analytical continuation $b_s^V (- {i\alpha \over 2k})$
 and $b_s^W (- {i\alpha \over 2k})$ in the formulas (51)--(53) draws attention.
 And what is the most essential, one
can see from the formula (53) that at any values of 
the momentum $k$ all three factors $v_i$ cannot
 vanish simultaneously. It means that the $\delta$-shaped additions (i.e.
 renormalization counterparts)
 are present {\it necessarily} in Eqs. (45), (47).

The substitution of the expressions (51), (52) in the formula (26) gives at 
$x \to 0^{+}$ the limiting value
$$
f_2 (k, 0^ +) =- {2ik \over {\alpha\, \Gamma \left( {i\alpha \over 2k}
\right)}}
    \, e^ {- {\pi \alpha \over 4k}} \,,                         \eqno {(54)}
$$
coinciding (that corresponds to the condition (38))
 with the limit $f_2 (k, 0^{-})$.
 Thus, we have
$$
|f_2(k, 0)|^2= {2k \over \pi \alpha} \, e^{- {\pi \alpha
    \over 2k}} \,
                 {\sinh \left({\pi\alpha \over 2k} \right)} \,  
                                                                 \eqno {(55)}
$$
at any set of admissible values of $v_i$, i.e. the functions $f_2 (k, x)$,
 satisfying
the system of Eqs. (38), (49), (50), are really well defined (as it was marked
above) and continuous in the singularity point.

The solutions (51)--(53) at imposing a number of additional conditions
describe a choice of permissible self-adjoint extensions. We shall consider
some of them.

{ \bf 1.} We shall search for values of $v_i$, at which the Schr{\"o}dinger
operator is real. We shall suppose in this case that the parameters $v_i$ are
real too.

The reality of the singular addition $\delta V_i (k, x)$ (46) on the
negative half-axis leads to the condition:
$$
v_-^{-, V} =0 \,.                                              \eqno {(56)}
$$
Using the resultant
from the formula (53) connection between $v_{+}^{+,W}$ and 
$v_{+}^{-,V}$ and imposing the reality condition for the singular
addition (48) on the positive half-line we obtain
 that at values of parameters (56) and
$$
v_ + ^ {-, V} = {-\pi + 2\, {\arg 2k} +
          \pi\, {\coth \left( {\pi\alpha \over 2k} \right)} \over
    - 2\, {\arg 2k} - \pi\, {\coth \left( {\pi\alpha \over 2k} \right)}} \,,
                                                              \eqno {(57)}
$$
$$
v_ + ^ {+, W} = {2 \left(\pi + {\arg 2k} -
      e^ {\pi \alpha \over k} \, {\arg 2k} \right)
          \over
      \left(-1 + e^ {\pi \alpha \over k} \right)
      \left(\pi + 2 {\arg 2k} \right)} \,,                     \eqno {(58)}
$$
the singular addition (48) is real and takes the form
$$
\delta V_ {f_2} (k, x) = -2 \alpha \left[2 \gamma_E +
      {\ln \left(2 {\sqrt {k^2}} x \right)} +
      {\rm Re} \, \psi \left({i \alpha \over 2k} \right) \right]
                \theta (x) \, \delta (x)                       \eqno {(59)}
$$
(here $\gamma_E$ is the  Euler constant; as
well as earlier, the equality $x\,\delta (x) =0$ is taken into
consideration calculating the formula (59)).

It is meaningful to pay attention to some properties of the addition (59). At
once one can see that the
addition (59) vanishes at switching off the interaction ($\alpha = 0$). 
Besides, the common sign ``minus'' in the
expression (59) shows that the correction to the Coulomb potential $\delta V_
{f_2} (k, x)$ weakens, ``compensates'' the initial interaction
 $\alpha | x |^{-1}$.
Using the representation for 
${\rm Re} \, \psi \left( {i \alpha \over 2k} \right)$
in the series (a35) it is possible to rewrite the expression (59) in the
form
$$
\delta V_ {f_2} (k, x) = -2 \alpha \left[ \gamma_E +
      {\ln \left( 2 {\sqrt {k^2}} x \right)} +
     {\alpha^2 \over 4k^2} \sum_ {n=1}^\infty 
                  {1 \over n} \, {1 \over {n^2 + {\alpha^2 \over 4k^2}}}
               \right] \theta(x) \, \delta (x) $$
   $$ \hphantom{aaaaa} \qquad \qquad \qquad \qquad \qquad \qquad 
                  (-\infty < {\alpha \over 2k} < \infty) \,
                                                   \qquad        \eqno {(60)}
$$
which is similar to the Feynman expansion.

The appropriate  to the low-energy case
 asymptotic expansion of ${\rm Re} \, \psi \left( {i \alpha \over 2k} \right)$
 is of interest also. The use of the 
expansion (a36)
gives for $\delta V_ {f_2} (k, x)$ the expression:
$$
\delta V_ {f_2} (k, x) \sim -2 \alpha \left[ 2\gamma_E +
      {\ln \left( \alpha x \right)} +
     \sum_ {n=1}^\infty 
                  {(-1)^{n-1} B_ {2n} \over {2n \left( \alpha \over 2k
\right)^{2n}}}
               \right] \theta (x) \, \delta (x) \quad (\alpha \gg 2k) \,
                                                                 \eqno {(61)}
$$
(here $B_ {2n}$ are the Bernoulli numbers).

For the case of scattering $\arg 2k = 0$, and
substituting the expressions (56)--(58) in Eqs. (51) and (52) we have
$$
{ \cal A} _R^ {(s)} (k) =- {2 \pi k \over \alpha} \,
 {1 \over \Gamma^2 \left({i\alpha \over 2k} \right)} \,
                 {\coth \left({\pi\alpha \over 2k} \right)} \,, \eqno {(62)}
$$
$$
{ \cal B} _R^ {(s)} (k) = - {\sinh \left({\pi\alpha \over 2k} \right)} \, 
                                                               , \eqno {(63)}
$$
hence
$$
 T_R (k) =- {\alpha \over 2 \pi k} \,
       \Gamma^2 \left({i\alpha \over 2k} \right) \,
                 {\tanh \left({\pi\alpha \over 2k} \right)} \,, \eqno {(64)}
$$
$$
 R_R (k) = {\alpha \over 2 \pi k} \,
       \Gamma^2 \left({i\alpha \over 2k} \right) \,
                 {\tanh \left({\pi\alpha \over 2k} \right)} \,
              {\sinh \left({\pi\alpha \over 2k} \right)} \,,    \eqno {(65)}
$$
i.e.
$$
{ \left| T_R (k) \right|} ^2 = {\rm sech} ^2 \left({\pi\alpha \over 2k}
\right) \,
                                                               , \eqno {(66)}
$$
$$
{ \left| R_R (k) \right|}^2={\tanh^2 \left({\pi\alpha \over 2k} \right)} \,
                                                               , \eqno {(67)}
$$
what ensures the fulfillment of the unitarity condition (43).

For bound states the values of the momentum located on the positive
imaginary axis correspond, i.e. it is necessary 
to put $\arg 2k=\pi/2$. In this case
$$
{ \cal A} _R^ {(b)} (k) = {2 \pi k \over \alpha} \,
 {1 \over \Gamma^2 \left({i\alpha \over 2k} \right)} \,
               e^ {\pi \alpha \over 2k} \,
                 {\rm cosech} \left({\pi\alpha \over 2k} \right) \,
                                                               . \eqno {(68)}
$$
The bound state locations are defined by simple poles
 of the coefficient $T_R(k)$, i.e. by simple zeros of the quantity (68).
 As ${\bigl(\Gamma (z)\bigr)}^{-1}$ is an entire function
 possessing the simple zeros at $z=-n$ it
is  easy to see that the expression (68) has the simple zeros at
$$
{ i \alpha \over 2k} =-n \,,                                   \eqno {(69)}
$$
i.e. the bound states exist at $\alpha < 0$ and have the negative 
energy
$$
E=- {\alpha^2 \over 4n^2} \,.                                  \eqno {(70)}
$$

Thus, the considered case leads to the old-known Bohr expression
 for the Coulomb levels of
of energy. But in the case under consideration the potential is not 
precisely the Coulomb one, the energy levels
(70) take place at presence of the additional point interaction (59).
Moreover, the corrections (59) written down in the form (60), (61) are
extremely similar to the corrections, arising in QED
 at the account of such phenomena as the
renormalizations, polarization of vacuum, vacuum fluctuations etc. (and
moreover, usually  it is accepted 
to interpret the specified corrections in QED as the relativistic
ones (see, e.c., \cite {bd-78})). It is essential that the received expressions
(and, in particular, the formula (59)) are {\it exact results}
 received beyond the framework of any approximations.

{ \bf 2.} The case of the complete transmittance.

The condition (38) (being, as was mentioned above, the condition of even
wave function continuation when passing through the singularity) can be
rewritten as
$$
R_R (k) -T_R (k) \, e^ {- {\pi \alpha \over 2k}} =
     {\Gamma \left({i\alpha \over 2k} \right)
        \over \Gamma \left(- {i\alpha \over 2k} \right)} \,.   \eqno {(71)}
$$
If one supposes that
$$
R_R (k) =0 \,,                                                 \eqno {(72)}
$$
then
$$
T_R (k) =e^ {\pi \alpha \over 2k} \,
     {\Gamma \left({i\alpha \over 2k} \right)
        \over \Gamma \left(- {i\alpha \over 2k} \right)} \,,    \eqno {(73)}
$$
that is
$$
{ \left| T_R (k) \right|} ^2 = 
      e^ {\pi \alpha \over k} \,,                             \eqno {(74)}
$$
and the unitarity condition (43) at real values of the momentum $k$
 and $\alpha \ne 0$ is not valid. Thus,
 at any values of the parameters $v_i$ 
the one-dimensional Coulomb potential is not absolutely transparent.

{ \bf 3.} The case of the absolute impenetrability of the Coulomb potential
corresponds to the choice of
$$
T_R (k) =0 \,.                                                \eqno {(75)}
$$
In this case
$$
R_R (k) = {\Gamma \left({i\alpha \over 2k} \right)
        \over \Gamma \left(- {i\alpha \over 2k} \right)} \,,   \eqno {(76)}
$$
$$
{ \left| R_R (k) \right|} ^2=1 \,.                              \eqno {(77)}
$$
The values of the scattering coefficients (75) and (76) are realized in three
cases: at the relationship
$$
v_+^{-,V}=-{2 + \left(1 - e^ {\pi \alpha \over k} \right) v_+^ {+,W}
      \over 1 + e^ {\pi \alpha \over k}} \,,                    \eqno {(78)}
$$
or at the values
$$
v_ + ^ {+, W} =-1 \qquad \left(v_-^ {-, V} \ne 1 \right)       \eqno {(79)}
$$
or
$$
v_ + ^ {-, V} =-1 \qquad \left(v_-^ {-, V} \ne 1 \right) \,.    \eqno {(80)}
$$
The values of two (and actually, due to the relation (53), 
one) parameters
$v_i$ remain arbitrary ones, but it appears insignificant: 
in all three cases (78)--(80)
irrespective to the value of left parameters the coefficients in the singular
addition $\delta V_ {f_2} (k, x) $ (48) approach the
infinity on the positive half-axis. Thus, the case 
of the complete impenetrability of the Coulomb potential
can be, in principle, treated, though this consideration has a rather formal
character. This case corresponds to the solution of the 
Dirichlet boundary problem
(see, e.c., \cite {aghh-91}), i.e. to the division of the axis
 into two independent intervals. In this case, obviously, it is
reasonable to consider as a solution of Eq. (47) 
the renormalized wave function determined by the condition
$$
f_2 (k, x) = {\cal A} _R (k) \, f_2^ {(R)} (k, x) \,,          \eqno {(81)}
$$
i.e.
$$
f_2^ {(R)} (k, x) = \left[\psi_ + ^ {-, V} (k, x) + 
   {\Gamma \left({i\alpha \over 2k} \right)
        \over \Gamma \left(- {i\alpha \over 2k} \right)} \,
          \psi_ + ^ {+, W} (k, x) \right] \, \theta (x) \,,    \eqno {(82)}
$$
and the function $f_2^{(R)} (k, x)$ (82) satisfies
 the Dirichlet boundary condition
$$
f_2^ {(R)} (k, 0) =0 \,.                                       \eqno {(83)}
$$

\section{Conclusions}

Summarizing the above it is possible to state that the method offered in the
paper allows to construct in view of the analytical structure of the 
fundamental solutions the
three-parametrical family of solutions (self-adjoint extensions) completely
determining the scattering coefficients, wave functions and corrections to the
Coulomb potential, and also to calculate appropriate energy levels. In this
connection all specified values appear unequivocally bound, interdependent.
Besides, it turns out that the corrections to the potential should be {\it
necessarily} present at any allowable values of parameters.
Also the remarkable similarity of the terms (59), (60) to the 
counterparts received by means of the dimensional regularization method
 attracts attention (and, what's more, these terms do not require the parameter
 breaking the scale invariance  at the 
dimensional regularization (see, e.g.,
\cite {ind-86}, as well as \cite {bdmcl-80})).

Thus, it is possible to speak that
at the level of calculations, connected with the nonrelativistic
Schr{\"o}dinger equation, there are
the renormalizations and terms appropriate
to radiation corrections of quantum electrodynamics, and 
moreover these quantities arise as a display of the internal structure 
of the appropriate dynamic equation. It
is essential that these quantities can be calculated without
 resorting to methods of the perturbation theory.

Taken as a whole the proposed method indicates to the opportunity
 of interpretation
of a wide class of phenomena in the quantum theory as the result  of the
appropriate Hamilton operator extension up to the self-adjoint one, and in the
part, concerning the renormalization theory, it can be considered as
a generalization of the Bogoliubov, Parasiuk and Hepp method of
renormalizations.

We shall notice finally that  the solution, corresponding to an even 
continuation of wave functions over the singularity point, is received in the 
present paper. The questions, related to a problem of degeneration (and, in 
particular, with an possibility of an odd continuation of wave function over zero), 
with an possibility of the inclusion of the $\delta'$-terms into the Hamiltonian and
detailed analysis of a spectrum of bound states (including a problem of absence of 
a level with indefinitely large negative energy) were not treated in the present 
paper. These problems will be analysed in the next publication.

\section*{Acknowledgments}

	 In conclusion the author supposes as a pleasant debt 
to express his sincere gratitude to B.A. Arbuzov,
V.I. Savrin and V.E. Troitski for the extremely valuable help and 
useful discussions.

%\newpage
%\bigskip

\section*{Appendix}

%\appendix {\Large\bf  Appendix}
%\medskip

The mathematical apparatus used in the paper is largely scattered among various
sources. There is no also uniform enough standard in the description of the
equations with the singular points. Besides, it is not possible 
to find a number of the formulas used in
the given article in the known mathematical manuals.
Therefore we shall give a brief report of the mathematical relations
 used in the paper. The designations basically correspond to Refs. 
\cite {olv-90} and \cite {abr-79}.

The equation like (1), (2) comes out of the confluent hypergeometric equation
$$
z\, {d^2w (z) \over dz^2} + (c-z) \, {dw (z) \over dz} -a\, w (z) =0\,,
\qquad
          a, \, c, \, z \in {\mbox {\mnf C}}                   \eqno {(a1)}
$$
except of the term of the first derivative. In this case it takes the
form
$$
{ d^2W (z) \over dz^2} = \left({1 \over 4} - {p \over z} + {m^2-1/4 \over
z^2}
        \right) \, W (z) \,,                                   \eqno {(a2)}
$$
where $p=c/2-a$, $m=c/2-1/2$, $W (z) =e^ {-z/2} \, z^ {m + 1/2} \, w (z) $,
 and it is referred to as the Whittaker equation.

Eq. (a1) has the regular singular point  with indices $0$ and $1-c$ at
the origin and the irregular singular point of rank $1$ at the
infinity.

The solution of Eq. (a1) represented in the series form and
corresponding to the index $0$ at the point $z=0$ takes the form
$$
M (a, c, z) =\sum_ {s=0} ^ {\infty} {(a) _s \over (c) _s} \, {z^s \over s!}
\,,
          \qquad c \ne 0,\, -1,\, -2,\, \ldots                \eqno {(a3)}
$$
( $ (a)_s$, $ (c)_s$ are the Pochhammer symbols). This series converges at all
finite $z$ and defines the Kummer function. The function $M (a, c, z) $ in the
general case is  a meromorphic function of variable $c$  with poles at the points
$0, -1, -2, \ldots$ and is an entire function in $a$. 
The following solution corresponds to the second index $1-c$
at $z=0$:
$$
N (a, c, z) = z^ {1-c} \, M (1 + a-c,\, 2-c,\, z) \,, \qquad
           c \ne 2,\, 3,\, 4,\, \ldots\,.                      \eqno {(a4)}
$$

Defining the Wronskian of two functions in the usual way:
$$
{ \cal W} \left\{w_1 (z), \, w_2 (z) \right\} =
      w_1 (z) \, {w_2} ' (z) - {w_1} ' (z) \, w_2 (z) \,,       \eqno {(a5)}
$$
it is easy to show that the Wronski determinant of the 
functions (a3) and (a4) is
of the form
$$
{ \cal W} \left\{M (a, c, z), \, N (a, c, z) \right\} =
      (1-c) \, e^z\, z^ {-c} \,.                              \eqno {(a6)}
$$

Pair of solutions of Eq. (a1) corresponding to the irregular singular
point, which are  linearly independent at all values of parameters, are the
functions $U (a, c, z)$ and $V (a, c, z)$ having normal expansions:
$$
U (a, c, z) \sim z^ {-a} \sum_ {s=0} ^ {\infty} (-1) ^s\, 
     {(a)_s\, (1 + a-c)_s \over s!\, z^s} \,, \quad z \to \infty\,,
  \quad | \arg z | \le {3 \pi \over 2} -\delta\,,               \eqno {(a7)}
$$
$$
V (a, c, z) \sim e^z\, (-z) ^ {a-c} \sum_ {s=0} ^ {\infty}
     {(c-a) _s\, (1-a) _s \over s!\, z^s} \,, \quad z \to \infty\,,
  \quad | \arg (-z) | \le {\pi \over 2} -\delta\,,              \eqno {(a8)}
$$
where $\delta$ is any small positive constant.
 $U (a, c, z)$ and $V (a, c, z)$ are many-valued functions of $z$
 entire in $a$
and $c$. The functions $U(a, c, z)$ and $V(a, c, z)$ are connected between
themselves by the relationship:
$$
V (a, c, z) =e^z\, U (c-a, c, -z) \,.                           \eqno {(a9)}
$$
The Wronski determinant for the functions $U(a, c, z)$ and $V(a, c, z)$ has
the form
$$
{ \cal W} \left\{U (a, c, z), \, V (a, c, z) \right\} =
      e^{\varepsilon ({\rm Im} \, z) \, i \pi (c-a)} \, e^z\, z^ {-c} \,
                                                            , \eqno {(a10)}
$$
where $\varepsilon ({\rm Im} \, z) $ is defined by the formula (3).

The standard solutions of Eq. (a2) are:
$$
M_ {p, m} (z) =e^ {- {z \over 2}} \, z^ {m + {1 \over 2}} \,
         M\left( m-p + {1 \over 2},\, 2m + 1,\, z \right) \,,  \eqno {(a11)}
$$
$$
W_ {p, m} (z) =e^ {- {z \over 2}} \, z^ {m + {1 \over 2}} \,
         U\left(m-p + {1 \over 2},\, 2m + 1,\, z \right) \,    \eqno {(a12)}
$$
and are referred to as the Whittaker functions. 
Each solution (a11) and (a12) is a
 many-valued function of $z$. The main branches correspond to the domain
 $\arg z \in (-\pi, \pi] $. Asymptotic behavior of the solutions
 (a11) and (a12) is defined by expressions:
$$
M_ {p, m} (z) \sim z^ {m + {1 \over 2}} \,, \quad z \to 0\,,
        \quad 2m \ne -1, \, -2, \, -3, \, \ldots \,,         \eqno {(a13)}
$$
$$
W_ {p, m} (z) \sim e^ {- {z \over 2}} \, z^p\,, \quad z \to \infty, \quad 
            | \arg z | \le {3 \pi \over 2} -\delta\,.        \eqno {(a14)}
$$
Since the scattering problem is investigated in the paper,
it is preferable to treat the
solutions possessing the asymptotic behavior like (a14). Therefore as a
fundamental pair we shall use the solutions connected with the functions
$$
W_ {p, m} (z) =e^ {- {z \over 2}} \, z^ {m + {1 \over 2}} \,
         U\left( m-p + {1 \over 2},\, 2m + 1,\, z \right) \,,  \eqno {(a15)}
$$
$$
V_ {p, m} (z) =e^ {- {z \over 2}} \, z^ {m + {1 \over 2}} \,
         V\left( m-p + {1 \over 2},\, 2m + 1,\, z \right) \,.  \eqno {(a16)}
$$
From the expansions (a7) and (a8) and the formula (a9) one can see
 that at $z \to \infty$
$$
V_ {p, m} (z) \sim e^ {z \over 2} \, z^ {-p} \,.              \eqno {(a17)}
$$
From comparison of the Whittaker equation (a2) and 
the stationary one-dimensional
Schr{\"o}dinger equation (1) with the potential (2) it is seen that at
$$
p=- {i\alpha \over 2k} \,, \quad m= {1 \over 2} \,, \quad z=-2ikx \,
                                                                \eqno {(a18)}
$$
the fundamental pair of solutions of Eqs. (1), (2) corresponding to
the scattering problem
 at $x > 0$ takes the form of (4a), (4b). As it is seen from the expression (2),
the transition to the negative half-axis corresponds to the replacement
 $\alpha \to -\alpha$ at $x < 0$, which
leads to the pair of solutions (5a), (5b). Normalization factors written out in
the expressions (4a)--(5b) lead to the asymptotic expansions (6a)--(6d).

The substitution (a18) reducing the Whittaker equation (a2) to the form (1),
(2) corresponds to the parameter $c$ in the functions 
$U (a, c, z)$, $V (a, c, z)$
having an integer value ($c=2$). It results in the necessity to use
 the limiting forms
of the functions $M (a, c, z) $ and $N (a, c, z) $
 (obtained, for example, with the
help of the Frobenius method) for the description of behavior of $U(a, c, z)$
and $V (a, c, z) $ near the point $z=0$. The appropriate logarithmic expansion
of the function $U (a, c, z) $ at $c=r$ ($r \in {\mbox {\mnf Z}} $) has in
general the following form
$$
U (a, r, z) =\sum_ {s=1} ^ {r-1} (-1) ^ {s-1} \,\lambda_ {r, -s} \, {(s-1)!
\over z^s} +
\lambda_ {r, 0} \, M (a, r, z) \,\ln z + $$
    $$ \sum_ {s=0} ^ {\infty} \lambda_ {r, s} \,\mu_ {r, s} \, {z^s \over s!}
\,,
                                                                \eqno {(a19)}
$$
where
$$
\lambda_ {r, s} = {(-1) ^r\,\Gamma (a + s) \over \Gamma (a) \,\Gamma (1 + a-r)
\, (r + s-1)!}
                                                           \,, \eqno {(a20)}
$$
$$
\mu_ {r, s} =\psi (a + s) -\psi (1 + s) -\psi (r + s) \,, \eqno {(a21)}
$$
$\psi (z) =\Gamma' (z) / \Gamma (z) $ is the digamma-function. 
Substitution of the values (a18) gives
 for the fundamental pair of solutions (4a)
and (4b) and their derivatives in the second order terms in $z$ the
expansions
$$
\psi_ + ^ {+, W} (k, x) = {2k \over 
      \alpha\,\Gamma \left({i \alpha \over 2k} \right)} \, 
         e^ {\pi \alpha \over 4k}
   \Biggl\{-i + x \Biggl[i \alpha - 2i\alpha \gamma_E + k -
             i\alpha\, {\ln (-2ik)} - $$
      $$ \left. \left. i\alpha\,\psi
 \left(1 + {i \alpha \over 2k} \right) - i\alpha\, {\ln x}
        \right] \right\} + {\rm O} (x^2) \,,                  \eqno {(a22)}
$$
$$
\left [\psi_ + ^ {+, W} (k, x) \right]' = {2k \over 
      \alpha\,\Gamma \left({i \alpha \over 2k} \right)} \, 
         e^ {\pi \alpha \over 4k}
   \left\{-2i\alpha \gamma_E + k - i\alpha\, {\ln (-2ik)} -
             i\alpha\,\psi \left(1 + {i \alpha \over 2k} \right) -
                                         \right. $$
    $$ i \alpha\, {\ln x} +
   x \left[2i \alpha^2 - 2i\alpha^2 \gamma_E + 3 \alpha k + ik^2 -
             i\alpha^2\, {\ln (-2ik)} + 2\alpha k\,
              \psi \left (1 + {i \alpha \over 2k} \right) - \right. $$
      $$ \left. \left. 
            i\alpha^2\,\psi \left (2 + {i \alpha \over 2k} \right) - 
            2\alpha k\,\psi \left (2 + {i \alpha \over 2k} \right) - 
                    i\alpha^2\, {\ln x}
        \right] \right\} + {\rm O} (x^2) \,,                   \eqno {(a23)}
$$
$$
\psi_ + ^ {-, V} (k, x) =- {2k \over 
      \alpha\,\Gamma \left(- {{i \alpha \over 2k}} \right)} \, 
         e^ {\pi \alpha \over 4k}
   \Biggl\{-i + x \Biggl[i \alpha - 2i\alpha \gamma_E - k -
             i\alpha\, {\ln (2ik)} - $$
      $$ \left. \left. i\alpha\,\psi
 \left(1- {i \alpha \over 2k} \right) - i\alpha\, {\ln x}
        \right] \right\} + {\rm O} (x^2) \,,                   \eqno {(a24)}
$$
$$
\left [\psi_ + ^ {-, V} (k, x) \right]' =- {2k \over 
      \alpha\,\Gamma \left(- {{i \alpha \over 2k}} \right)} \, 
         e^ {\pi \alpha \over 4k}
   \left\{-2i\alpha \gamma_E - k - i\alpha\, {\ln (2ik)} -
             i\alpha\,\psi \left(1- {i \alpha \over 2k} \right) -
                                          \right. $$
    $$ i\alpha\, {\ln x} + 
  x \left[2i \alpha^2 - 2i\alpha^2 \gamma_E - 3 \alpha k + ik^2 -
             i\alpha^2\, {\ln (2ik)} - 2\alpha k\,
              \psi \left(1- {i \alpha \over 2k} \right) - \right. $$
      $$ \left. \left. 
            i\alpha^2\,\psi \left(2- {i \alpha \over 2k} \right) + 
            2\alpha k\,\psi \left(2- {i \alpha \over 2k} \right) - 
                    i\alpha^2\, {\ln x}
        \right] \right\} + {\rm O} (x^2) \,                   \eqno {(a25)}
$$
( Here $\gamma_E$ is the Euler constant). As it was 
specified above the replacement
$\alpha \to -\alpha$ gives appropriate expansions for the fundamental pair (5a),
(5b).

Earlier it was already noted that $U(a, c, z)$ and $V(a, c, z)$ are
many-valued functions of $z$. Using relations
$$
W_ {p, m} (z) = {\Gamma (-2m) \over \Gamma (1/2-m-p)} \, M_ {p, m} (z) +
       {\Gamma (2m) \over \Gamma (1/2 + m-p)} \, M_ {p,-m} (z) \,,
                                                              \eqno{(a26)}
$$
$$
V_ {p, m} (z) = {\Gamma (-2m) \over \Gamma (1/2-m + p)} \, M_ {p, m} (z) +
   {\Gamma (2m)\,e^ {-2i \pi m} \over \Gamma (1/2 + m + p)}\,M_ {p,-m}(z)
                                                          \,, \eqno {(a27)}
$$
and also taking into account that
$$
M_ {-p, m} \left(z\, e^ {i \pi} \right) = i\, e^ {i \pi m} \, M_ {p, m} (z)
                                                         \,, \eqno {(a28)}
$$
it is possible to get the formulas for the analytical continuations
 (at $s \in {\mbox {\mnf Z}}$)
$$
(-1)^s\, W_ {p, m} \left(z\, e^ {2i \pi s} \right) =
- {{e^ {2i \pi p} \, {\sin (2 \pi sm)} + {\sin [2 \pi (s-1) m]}} \over 
       \sin (2 \pi m)} \, W_ {p, m} (z) + $$
     $$
   {\sin (2 \pi sm) \over \sin (2 \pi m)} \,
   {2 \pi\,e^{i \pi (p + m)} \over \Gamma (1/2 + m-p)\,\Gamma (1/2-m-p)} \,
            V_ {p, m} (z) \,,                                 \eqno {(a29)}
$$
$$
(-1)^s\, V_ {p, m} \left(z\, e^ {2i \pi s} \right) =
- {{e^ {2i \pi p} \, {\sin (2 \pi sm)} + {\sin [2 \pi (s + 1) m]}} \over 
       \sin (2 \pi m)} \, V_ {p, m} (z) - $$
     $$
   {\sin (2 \pi sm) \over \sin (2 \pi m)} \,
   {2 \pi\,e^{i \pi (p-m)} \over \Gamma (1/2-m + p)\,\Gamma (1/2 + m + p)}\,
            W_ {p, m} (z) \, .                               \eqno {(a30)}
$$
In the considered case it is natural to use the limiting forms of expressions
for $U (a, c, z) $ and $V (a, c, z) $.

Except of the Wronskians (7), (8) mentioned in the basic text
the following relations are used in the paper:
$$
{ \cal W} \left\{\left[\psi_ + ^ {-, V} (k, x) \right]^*, \,
              \psi_ + ^ {-, V} (k, x) \right\} = -2ik \,,      \eqno {(a31)}
$$
$$
{ \cal W} \left\{\left[\psi_ + ^ {+, W} (k, x) \right]^*, \,
              \psi_ + ^ {+, W} (k, x) \right\} = 2ik \,,       \eqno {(a32)}
$$
$$
{ \cal W} \left\{\left[\psi_ + ^ {-, V} (k, x) \right]^*, \,
              \psi_ + ^ {+, W} (k, x) \right\} = 0 \,,         \eqno {(a33)}
$$
and moreover
$$
 \left[\psi_ + ^ {-, V} (k, x) \right]^* = \psi_ + ^ {+, W} (k, x) \,.
                                                                \eqno{(a34)}
$$

At the analysis of the expression (59)
the series expansion of the function ${\rm Re} \,\psi (z)$ 
is used in the paper. At $z=x + iy$
$$
{ \rm Re} \, \psi (iy) = {\rm Re} \, \psi (-iy) = 
{ \rm Re} \, \psi (1 + iy) = {\rm Re} \, \psi (1-iy) = $$
       $$ - \gamma_E + y^2 \sum_ {n=1} ^\infty {1 \over n} \, {1 \over {n^2 +
         y^2}}  \qquad   (- \infty < y < \infty) \,. \         \eqno {(a35)}
$$
The asymptotic formulas for ${\rm Re} \, \psi (iy)$ has the form
$$
{ \rm Re} \, \psi (iy) \sim
      \ln y + \sum_ {n=1}^\infty 
                  {(-1)^{n-1} B_ {2n} \over 2n \, y^ {2n}} = $$
        $$ \ln y + {1 \over 12y^2} + {1 \over 120y^4} + {1 \over 252y^6} +
\ldots     \qquad       (y \to \infty)                         \eqno {(a36)}
$$
($B_ {2n} $ are the Bernoulli numbers).

\end {document}